\documentclass[twocolumn,aps,prl,superscriptaddress,floatfix]{revtex4}
\usepackage{color}
\usepackage{amsmath}
\usepackage{amssymb}
\usepackage{graphicx}

\makeatletter
\usepackage{epsfig}
\usepackage{color}
\usepackage{ulem}

\begin{document}
\title{Probing Collective and Individual Kondo Screening: \\Multi-Stage, Multi-Channel Kondo Effects in a $C_3$-Symmetric Four-Impurity Model}
\author{Danqing Hu}
\email[]{hudq@cqu.edu.cn}
\affiliation{Department of Physics and Chongqing Key Laboratory for Strongly Coupled Physics, Chongqing University, Chongqing 401331, China}
\author{Jiangfan Wang}
\affiliation{School of Physics, Hangzhou Normal University, Hangzhou, Zhejiang 311121, China}
\author{Zixiang Hu}
\email[]{zxhu@cqu.edu.cn}
\affiliation{Department of Physics and Chongqing Key Laboratory for Strongly Coupled Physics, Chongqing University, Chongqing 401331, China}
\author{Yi-feng Yang}
\email[]{yifeng@iphy.ac.cn}
\affiliation{Beijing National Laboratory for Condensed Matter Physics and Institute of Physics, Chinese Academy of Sciences, Beijing 100190, China}
\affiliation{School of Physical Sciences, University of Chinese Academy of Sciences, Beijing 100049, China}

\date{\today}

\begin{abstract}
Collective screening has been proposed to underlie the basic physics of multi-impurity and lattice Kondo systems. But how to establish this picture and distinguish it from individual (local) Kondo screening remains a grand challenge. The recently developed auxiliary-bath numerical renormalization group (AUNRG) method provides a key step towards resolving this issue. Its application to the $C_3$-symmetric three-impurity Kondo (3IK) model with a shared electron bath reveals fully screened Fermi liquid ground states arising from collective screening of cluster spin degrees of freedom at small Kondo coupling $J_{\rm K}$ and individual (local) screening of local impurities at large $J_{\rm K}$. Here we design a four-impurity Kondo (4IK) model where the probe impurity couples only to the original three Kondo impurities to detect the nature of the screened states. We show that the collective and individual Kondo screenings give rise to emergent multi-stage, two-channel Kondo effect and unstable three-channel Kondo effect, respectively. This suggests a special helicity structure of the screened states and confirms the idea of collective screening in multi-impurity Kondo systems. Our work also demonstrates that the same auxiliary-bath construction can be readily extended to models with additional impurities to explore novel many-body quantum states with emergent cluster degrees of freedom.
\end{abstract}

\maketitle
\textit{Introduction.--}
Kondo impurity models are often regarded as the simplest correlated electron systems \cite{Kondo1964,Anderson1961,Nozieres1974,Hewson1993}. In combination with the dynamical mean-field theory, the single-impurity Kondo problem has helped to shape our understanding of correlated electron materials such as transition metal oxides and rare-earth or actinide intermetallics \cite{Georges1996,Maier2005}. By contrast, the multi-impurity Kondo problems have remained a grand challenge, thus hindering a deeper understanding of real materials beyond local approximations \cite{Jones1987,Jones1988,Jayaprakash1981,Paul1996,Ingersent2005,Zitko2007,Mitchell2011,Mitchell2012,Mitchell2013,Liu2016PRB,Eickhoff2020,Wojcik2020PRB,Konig2021,ColemanNevidomskyy2010,Wang2022PRB,Ferrero2007}. Recently, an auxiliary-bath numerical renormalization group (AUNRG) method has been developed for solving general multi-impurity models \cite{Hu2026AUNRG,Wang2024}. By mapping the shared conduction electron baths to several auxiliary baths, AUNRG provides a transparent and fully controlled method to construct independent Wilson chains, thus allowing for direct application of mature NRG algorithms \cite{Wilson1975,Yoshida1990,Oliveira1994,Bulla2008,Mitchell2014,Stadler2016} to multi-impurity Kondo models. Its first application to the $C_3$-symmetric three-impurity Kondo (3IK) models with shared conduction bath reveals collective Kondo screening of cluster degrees of freedom and excludes the speculated non-Fermi-liquid fixed point under independent parameter assumptions in realistic systems \cite{Paul1996,Ingersent2005,Eickhoff2020}. The success raises a number of additional questions: How can one probe the collective screening and distinguish it from the local screening of individual Kondo impurities? Could the three-impurity models be modified to yield other exotic quantum states? Can these states be manipulated by controlling the emergent degrees of freedom?

In this work, we propose a simple extension of the 3IK model to a $C_3$-symmetric four-impurity Kondo (4IK) model by introducing an additional probe impurity. We show that previous construction of the Wilson chains for the 3IK model can be readily extended to the 4IK model as long as the probe impurity is not coupled with the shared bath. Our AUNRG calculations then reveal rich emergent phenomena such as two-channel and three-channel Kondo screened states with the residual impurity entropy $S_{\rm imp}=\frac{1}{2}\ln 2$ and $\ln\frac{1+\sqrt{5}}{2}$, respectively, helicity-degenerate spinless state with residual entropy $S_{\rm imp}=\ln 2$, and collective Kondo screening of spin-1 cluster. In the two-channel Kondo regime, collective Kondo screening of the helicity $\pm1$, spin-1/2 cluster states in the original 3IK model first produces a helicity-degenerate Fermi liquid state, which then compete to screen the probe impurity, resulting in the overscreened two-channel Kondo fixed point \cite{Tsvelick1985,Zitko2007,Mitchell2011,Mitchell2012}. Likewise, the three-channel Kondo physics \cite{Tsvelick1985,Affleck1993,Stadler2016} arises when the three Kondo impurities are individually screened and produce three (slightly-mixed) local Fermi liquids that compete to screen the probe impurity. Thus, the two-channel and three-channel physics reflect exactly the predicted collective and individual Kondo screening of the original 3IK model. Our work greatly extends the application scope of the AUNRG to models with additional impurities, provides supporting evidence for collective screening of the cluster spin/helicity states, and opens the avenue for exploring more correlated phenomena by manipulating the emergent cluster degrees of freedom.

\begin{figure}
\centering
\includegraphics[width=0.48\textwidth]{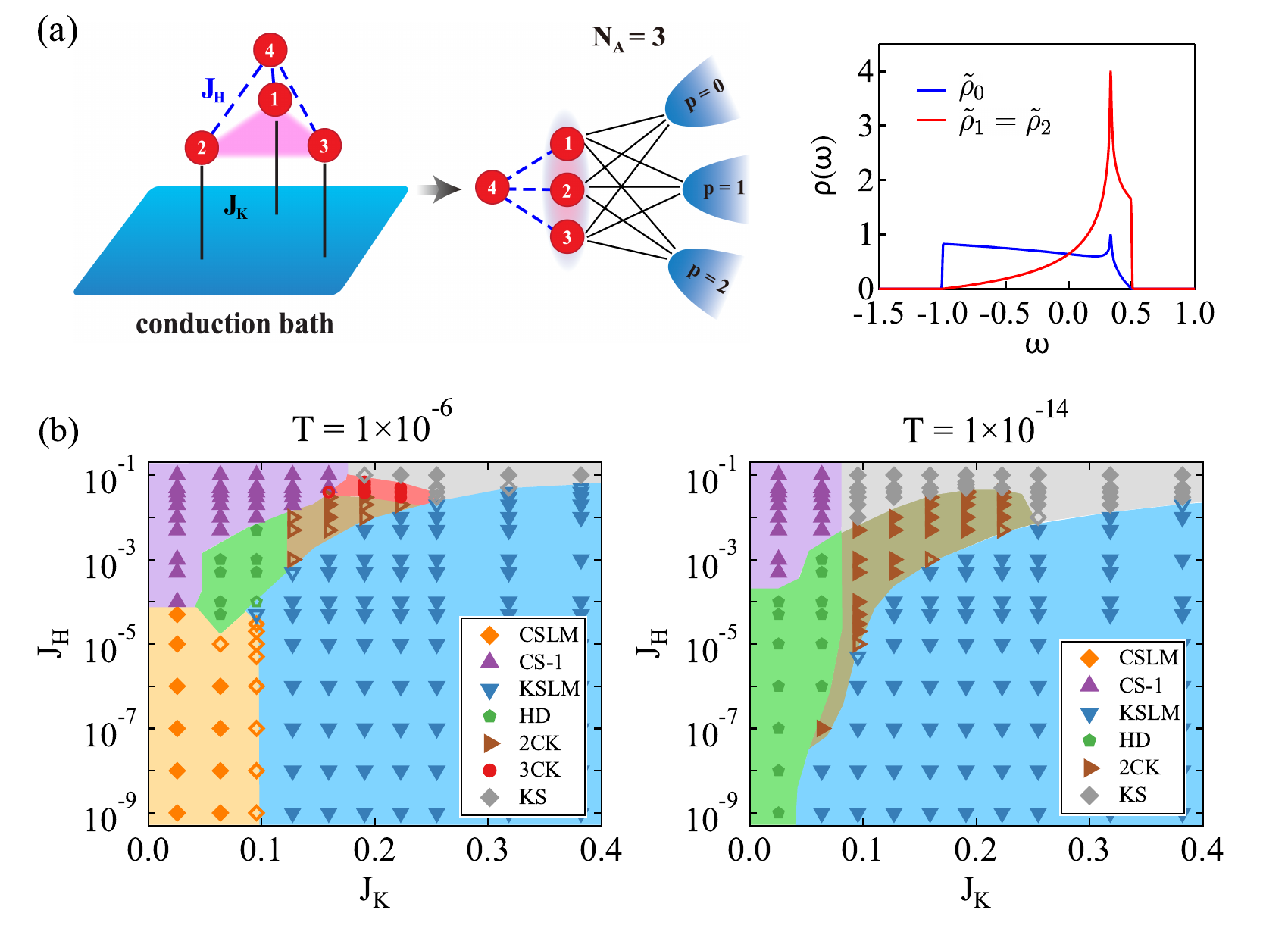}
\caption{(a) Illustration of the proposed 4IK model with three Kondo impurities ($\mu=1$, 2, 3) coupled ($J_{\rm K}$) symmetrically to a single shared bath and an additional probe impurity ($\mu=4$) only coupled ($J_{\rm H}$) to the Kondo impurities. The shared bath is then mapped to three auxiliary baths labeled by $p=0$, 1, 2, whose densities of states $\tilde{\rho}_p$ are also plotted for clarity. (b) Numerical phase diagrams on the $J_{\rm K}-J_{\rm H}$ plane at two typical temperatures $T=1\times 10^{-6}$ and $T=1\times 10^{-14}$, showing emergent quantum many-body states (filled symbols); open symbols mark the crossover region between two phases. CSLM denotes the cluster spin-1/2 state of three Kondo impurities (CS-1/2) weakly coupled to the probe impurity; KSLM denotes the fully screened state of three Kondo impurities weakly coupled to the probe impurity; CS-1 denotes the global spin-1 cluster state of all four impurities; HD denotes the spinless helicity-degenerate cluster state; 2CK and 3CK denote the two-channel and three-channel Kondo states, respectively; and KS marks the different fully screened Kondo states.}
\label{fig1}
\end{figure}

\textit{Model and method.--}
We start with the Hamiltonian,
\begin{align}
H=\sum_{\boldsymbol{k}\sigma}\epsilon_{\boldsymbol{k}}c_{\boldsymbol{k}\sigma}^{\dagger}c_{\boldsymbol{k}\sigma}+J_{\rm K}\sum_{\mu=1}^{3}\boldsymbol{s}_{\mu}\cdot\boldsymbol{S}_{\mu}+J_{\rm H}\sum_{\mu=1}^{3}\boldsymbol{S}_{4}\cdot\boldsymbol{S}_{\mu},
\label{eq:Hmic}
\end{align}
where $\epsilon_{\bf k}$ describes the shared conduction bath, $\boldsymbol{s}_{\mu}$ ($\mu=1,\ 2,\ 3$) represents the spin density operator at ${\bf r}_\mu$, and $J_{\rm K}$ denotes their local coupling to the three Kondo impurities $\boldsymbol{S}_{\mu}$, which couple equally to the probe impurity $\boldsymbol{S}_{4}$ via $J_{\rm H}\ge0$. Following the same construction for the 3IK model \cite{Hu2026AUNRG,Wang2024}, the conduction bath is disentangled into three auxiliary baths denoted by $p=0,\ 1,\ 2$ with the densities of states (DOS), $\tilde{\rho}_0(\omega)=\rho_0(\omega)+2\rho_1(\omega)$ and $\tilde{\rho}_1(\omega)=\tilde{\rho}_2(\omega)=\rho_0(\omega)-\rho_1(\omega)$, which are obtained by diagonalizing the $3\times3$ matrix of the original densities of states (DOS), $\rho_{\mu\nu}(\omega)\equiv\sum_{\boldsymbol{k}}e^{i\boldsymbol{k}\cdot(\boldsymbol{r}_{\mu}-\boldsymbol{r}_{\nu})}\delta(\omega-\epsilon_{\boldsymbol{k}})$, with the diagonal (local) elements $\rho_{\mu\mu}(\omega)=\rho_0(\omega)$ and the off-diagonal (nonlocal) elements $\rho_{\mu\nu}(\omega)=\rho_1(\omega)$ for $\mu\neq\nu$. Correspondingly, the spin density operators are mapped to $\boldsymbol{s}_{\mu}\rightarrow \sum_{pp'}w^*_{\mu p}w_{\mu p'}\boldsymbol{s}_{pp'}$, via the $3\times3$ transformation matrix $w_{\mu p}=\frac{1}{\sqrt{3}}e^{-i2\pi(\mu-1)p/3}$. Obviously, including the probe impurity does not affect $\rho_{\mu\nu}(\omega)$ and its diagonalization, demonstrating that the same auxiliary-bath construction can be readily applied to extended models with additional impurities decoupled from the bath.

For clarity, we present the model geometry and the auxiliary DOS in Fig.~\ref{fig1}(a). Only the $\tilde{\rho}_p(\omega)$ are needed for constructing the Wilson chains. NRG calculations are then performed with $N_s=4000$ kept states and the discretization parameter $\Lambda=10$, and the results are examined for other choices of parameters \cite{Wilson1975,Bulla2008,Yoshida1990,Oliveira1994}. All presented results are obtained after averaging over $z=0$ and 1/2. The conduction bandwidth is set to $D=1.5$ as the energy unit. The different emergent states are distinguished by their distinct impurity entropy $S_{{\rm imp}}=S-S_0$, spin correlation functions $\langle\boldsymbol{S}_1\cdot\boldsymbol{S}_2\rangle$ and $\langle\boldsymbol{S}_1\cdot\boldsymbol{S}_4\rangle$, and effective moment $\mu_{\rm eff}=T\chi=\langle \boldsymbol{S}_z^2\rangle - \langle \boldsymbol{S}_z^2 \rangle_0$, where $S$ is the total entropy, $\boldsymbol{S}_z$ is the $z$-component of the total spin $\boldsymbol{S}$, and the subscript $0$ refers to the noninteracting bath only. For simplicity, the impurity $g$ factor, the Bohr magneton $\mu_B$, and the Boltzmann constant $k_B$ are all set to unity.

\begin{table}[t]
\caption{\label{tab1}Signatures of major emergent states of the proposed 4IK model, characterized by the impurity entropy $S_{\rm imp}$, the effective moment $\mu_{\rm eff}$, and the spin correlations $\langle\boldsymbol{S}_{1}\cdot\boldsymbol{S}_{2}\rangle$ and $\langle\boldsymbol{S}_{1}\cdot\boldsymbol{S}_{4}\rangle$. The 3CK and CS-1 states are unstable and only exist at finite temperatures.}
\begin{ruledtabular}
\begin{tabular}{lcccc}
Emergent state & $S_{\rm imp}$ & $\mu_{\rm eff}$ & $\langle\boldsymbol{S}_{1}\cdot\boldsymbol{S}_{2}\rangle$ & $\langle\boldsymbol{S}_{1}\cdot\boldsymbol{S}_{4}\rangle$\\
\hline
CSLM & $3\ln 2$ & $1/4 + 1/4$ & $-1/4$ & $0$\\
KSLM & $\ln 2$ & $1/4$ & $-1/4$ & weak AFM\\
CS-1 & $\ln 3$ & $\sim2/3$ & $+1/4$ & AFM\\
HD & $\ln 2$ & $0$ & $-1/4$ & $-1/4$\\
2CK & $\frac{1}{2}\ln 2$ & $0$ & $-1/4$ & weak AFM\\
3CK & $\ln\frac{1+\sqrt{5}}{2}$ & $0$ & $ 0$ & weak AFM\\
KS & 0 & 0 & - & -\\
\end{tabular}
\end{ruledtabular}
\end{table}

\textit{Global phase diagram.--}
The major emergent states are presented on the $J_{\rm K}$-$J_{\rm H}$ phase diagrams at two characteristic temperatures $T=1\times 10^{-6}$ and $T=1\times 10^{-14}$ in Fig.~\ref{fig1}(b). For clarity, the characteristic signatures of each state are listed in Table~\ref{tab1}, which can be roughly categorized into two classes depending on the strength of $J_{\rm H}$ relative to the collective or individual screening temperatures of the three Kondo impurities. In the original 3IK model, the Kondo impurities form a helicity $\pm1$, spin-1/2 cluster state (CS-1/2) at small $J_{\rm K}$ when the RKKY interaction dominates, which is then collectively screened to yield a Fermi liquid ground state at low temperatures \cite{Hu2026AUNRG}. For large $J_{\rm K}$ where the Kondo effect dominates, the impurities are individually screened and form (slightly-mixed) local screened states at three impurity sites \cite{Hu2026AUNRG}. The effect of $J_{\rm H}$ can be understood from these starting points.

For small $J_{\rm H}$, the probe impurity couples weakly to the screened states. At small $J_{\rm K}$, the screened state consists of two helicity channels that compete to screen the probe impurity, resulting in a second-stage two-channel Kondo (2CK) screening. Likewise, at large $J_{\rm K}$, the three local Kondo screened states compete to screen the probe impurity, causing a second-stage three-channel Kondo (3CK) screening, which is eventually taken over by a fully screened Fermi liquid ground state due to the mixture of the three local screened states. The 2CK and 3CK effects probe respectively the collective and individual Kondo screening of the original 3IK model.

When $J_{\rm H}$ exceeds the screening temperatures, the probe impurity first couples to the Kondo impurities and forms cluster spin states. At small $J_{\rm K}$, this leads to a global spin singlet between the probe impurity and the helicity $\pm1$, spin-1/2 cluster, resulting in a spinless helicity-degenerate cluster state (HD). At large $J_{\rm K}$, where the RKKY interaction is relatively small, the probe impurity may force all three Kondo impurities to align towards its antiparallel direction, thus producing a global spin-1 cluster state (CS-1).

Below we provide detailed numerical evidence for these emergent states, first discussing briefly the HD and CS-1 states and then focusing on the 2CK and 3CK regimes.

\begin{figure}
\centering
\includegraphics[width=0.48\textwidth]{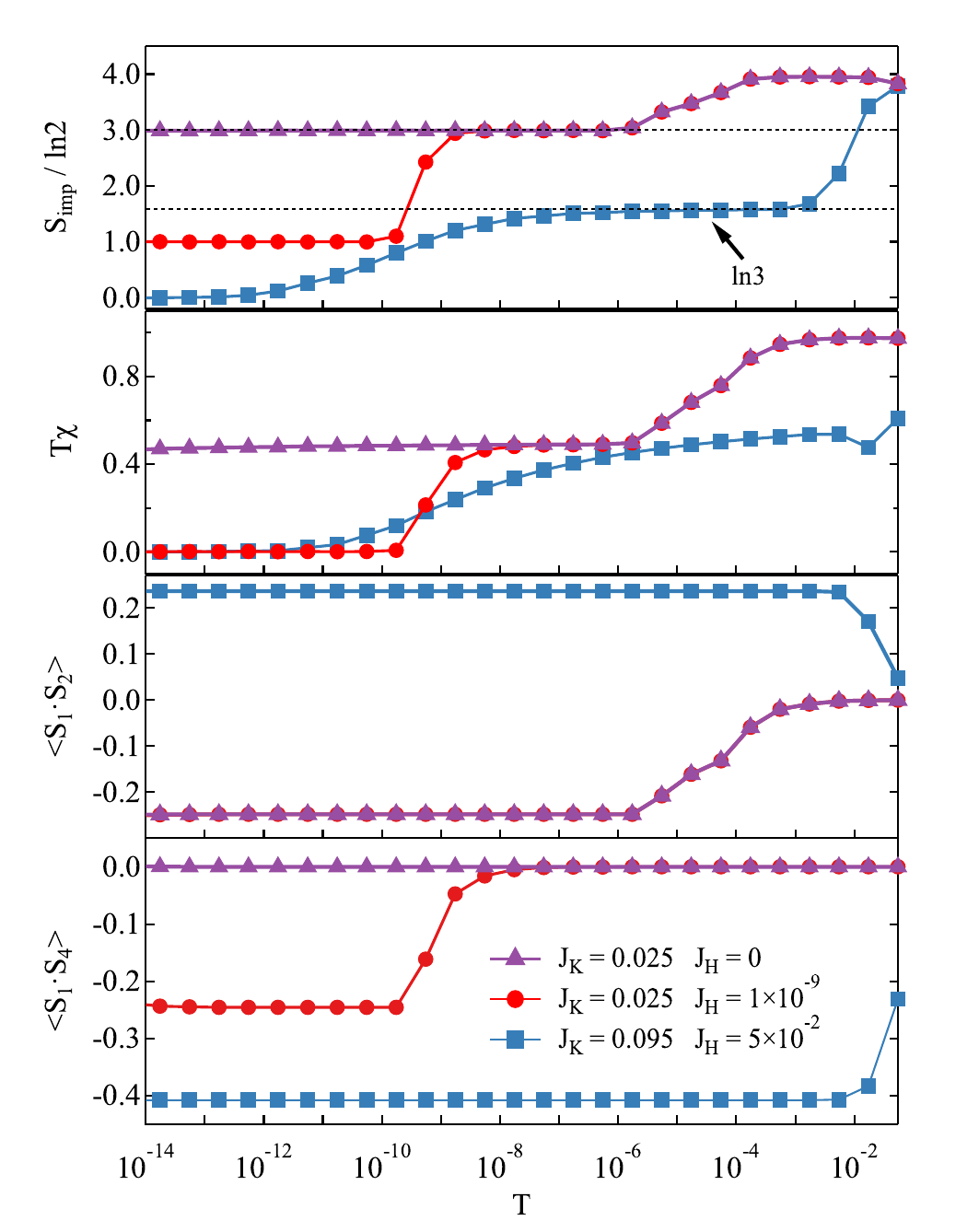}
\caption{Emergence of the HD and CS-1 cluster states from the temperature evolution of the impurity entropy $S_{\rm imp}$, the effective moment $\mu_{\rm eff}\equiv T\chi$, and the spin correlations $\langle\boldsymbol{S}_1\cdot\boldsymbol{S}_2\rangle$ and $\langle\boldsymbol{S}_1\cdot\boldsymbol{S}_4\rangle$ for $J_{\rm K}=0.025,\,J_{\rm H}=1\times10^{-9}$ and $J_{\rm K}=0.095,\,J_{\rm H}=5\times10^{-2}$. The $J_{\rm K}=0.025, \,J_{\rm H}=0$ results are also shown for reference.}
\label{fig2}
\end{figure}

\textit{Global helicity/spin states.--}
Figure~\ref{fig2} shows an example of the HD state at $J_{\rm K}=0.025$ and $J_{\rm H}=1\times10^{-9}$ (red circles). Compared to the results for $J_{\rm H}=0$, the impurity entropy drops from $3\ln 2$ for the CS-1/2 state plus free spin-1/2 probe impurity to $\ln 2$ at $T\approx J_{\rm H}$, where the effective moment also reduces from approximately $1/4+1/4=1/2$ to zero, the probe impurity develops an antiferromagnetic spin correlation with the cluster spin, $\langle\boldsymbol{S}_1\cdot\boldsymbol{S}_4\rangle\approx-1/4$, while the internal correlation among Kondo impurities remains unchanged. This demonstrates that the probe impurity forms a global spin singlet with the CS-1/2 state and the spin degrees of freedom are fully quenched, leaving only the helicity degree of freedom that contributes the residual entropy $\ln2$.

An example of the CS-1 state is also given in Fig.~\ref{fig2} at $J_{\rm K}=0.095$ and $J_{\rm H}=5\times10^{-2}$ (blue squares). As the temperature decreases, the entropy first drops to $\ln (2\tilde{S}+1)=\ln 3$ with the effective moment approaching $\tilde{S}(\tilde{S}+1)/3\approx2/3$. In the meanwhile, the internal spin correlation among Kondo impurities becomes positive (ferromagnetic) and that with the probe impurity turns negative (antiferromagnetic), indicating that the large $J_{\rm H}$ forces all three Kondo impurities to align antiparallel with the probe impurity, forming a four-impurity cluster state of total spin-1. Upon further cooling, the impurity entropy and effective moment both drop to zero, indicating that this spin-1 moment is eventually screened by the conduction bath.

\begin{figure}[t]
\centering
\includegraphics[width=0.48\textwidth]{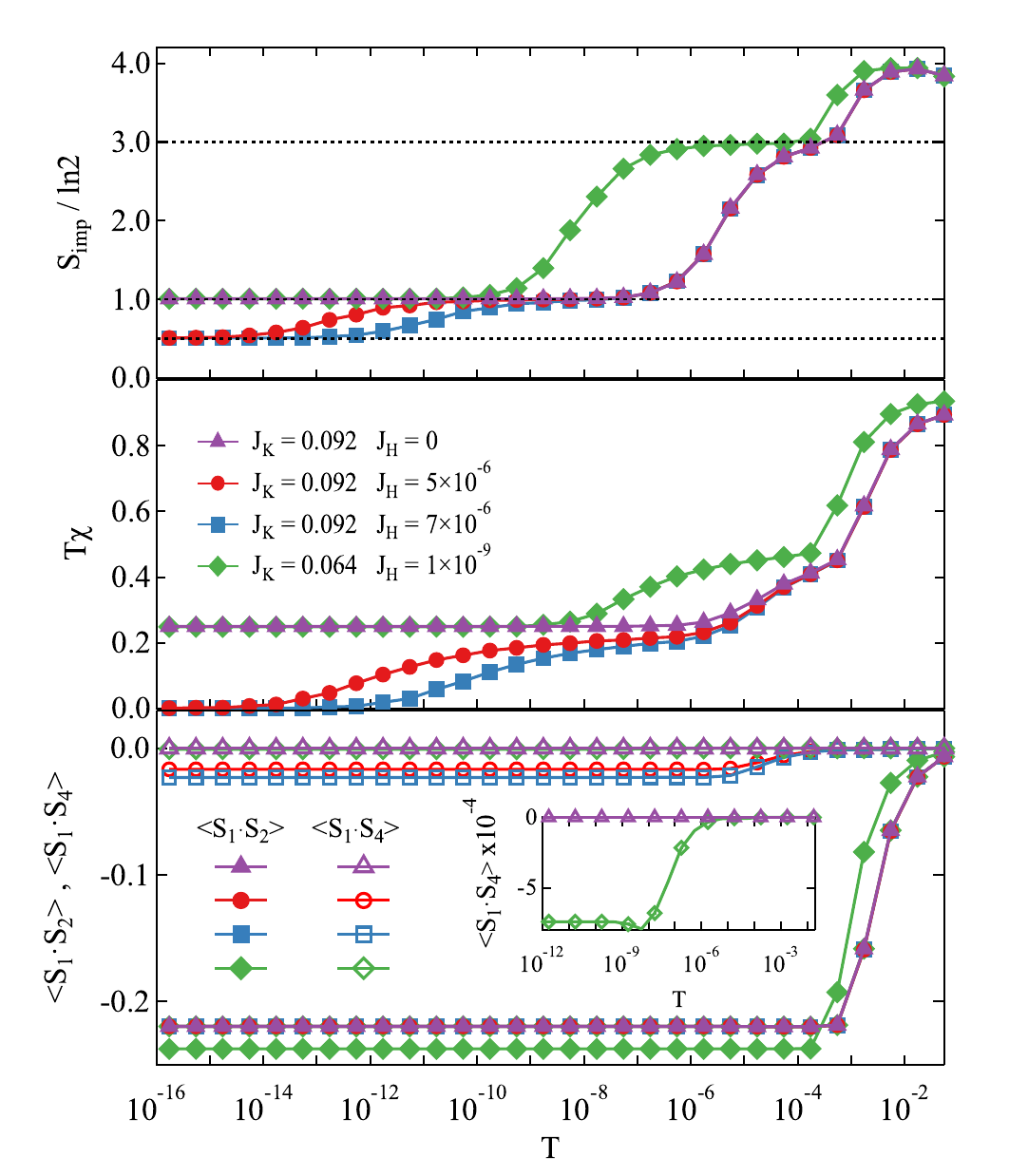}
\caption{Identification of the 2CK state from the temperature evolution of the impurity entropy $S_{\rm imp}$, the effective moment $\mu_{\rm eff}\equiv T\chi$, and the spin correlations $\langle\boldsymbol{S}_1\cdot\boldsymbol{S}_2\rangle$ (filled symbols) and $\langle\boldsymbol{S}_1\cdot\boldsymbol{S}_4\rangle$ (open symbols) for $J_{\rm H}=5\times10^{-6}$ and $7\times10^{-6}$ at $J_{\rm K}=0.092$ and $J_{\rm H}=1\times10^{-9}$ at $J_{\rm K}=0.064$. The 2CK state is characterized by the $\frac{1}{2}\ln 2$ entropy plateau. The inset shows the enlarged plot for the weak AFM correlation between the probe and Kondo impurities. The $J_{\rm H}=0$ results at $J_{\rm K}=0.092$ are also shown for reference.}
\label{fig3}
\end{figure}

\textit{Two-channel Kondo state.--}
Figure~\ref{fig3} shows two examples of the 2CK state with $J_{\rm H}=5\times10^{-6}$ and $7\times10^{-6}$ at $J_{\rm K}=0.092$. As expected, the impurity entropy first drops to $3\ln 2$, accompanied with a sudden change of $\langle\boldsymbol{S}_1\cdot\boldsymbol{S}_2\rangle$ from roughly zero to -0.22, whereas $\langle\boldsymbol{S}_1\cdot\boldsymbol{S}_4\rangle$ remains zero, indicating the onset of the CS-1/2 state with isolated probe impurity. This regime occurs in the RKKY dominant regime at small $J_{\rm H}$.

As the temperature is further lowered, the impurity entropy drops to $\ln 2$ and the effective moment reduces from 1/2 to approximately 1/4, while the spin correlations remain unchanged, reflecting collective screening of the helicity $\pm1$, spin-1/2 cluster (CS-1/2) whose internal spin correlation is not affected, while the entropy and effective moment are mainly contributed by the probe impurity. Further lowering the temperature reduces the entropy to a $\frac{1}{2}\ln 2$ plateau and $\mu_{\rm eff}$ to 0, but leaves the spin correlations unchanged. Comparison with the $J_{\rm H}=0$ results suggests that the probe impurity spin is also screened at this stage, and the residual $\frac{1}{2}\ln 2$ entropy signals a 2CK effect \cite{Tsvelick1985,Zitko2007,Mitchell2011,Mitchell2012}, which can only arise if the probe impurity is screened simultaneously by two symmetric helicity channels of the Kondo-screened cluster state indirectly through $J_{\rm H}$. In fact, as shown in Fig.~\ref{fig3}, increasing $J_{\rm H}$ enhances primarily the crossover temperature to the $S_{\rm imp}=\frac{1}{2}\ln 2$ plateau. By contrast, if $J_{\rm H}$ is too small, as shown for $J_{\rm H}=0$ or $J_{\rm K}=0.064$ and $J_{\rm H}=1\times10^{-9}$, the probe impurity remains unscreened down to the lowest temperature.

\begin{figure}
\centering
\includegraphics[width=0.48\textwidth]{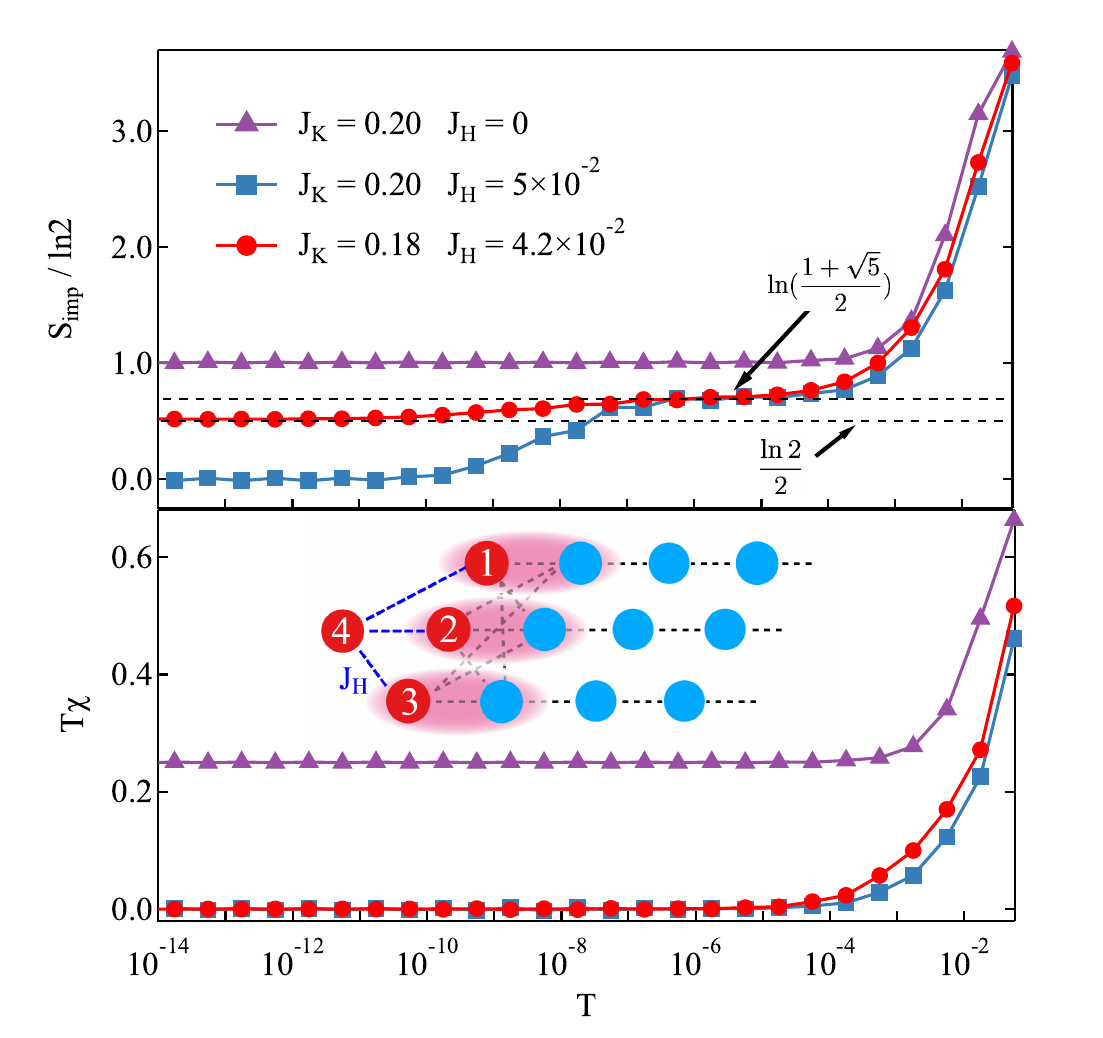}
\caption{Identification of the 3CK state from the temperature evolution of the impurity entropy $S_{\rm imp}$ and the effective moment $\mu_{\rm eff}\equiv T\chi$ for $J_{\rm K}=0.20$,  $J_{\rm H}=5\times10^{-2}$ and $J_{\rm K}=0.18$, $J_{\rm H}=4.2\times10^{-2}$. The 3CK state is unstable and characterized by the $\ln\frac{1+\sqrt{5}}{2}$ entropy plateau. The inset illustrates its formation. The $J_{\rm H}=0$ results at $J_{\rm K}=0.20$ are also shown for reference.}
\label{fig4}
\end{figure}

\textit{Three-channel Kondo state.--}
Figure~\ref{fig4} shows two examples of the 3CK state with $J_{\rm K}=0.20$, $J_{\rm H}=5\times10^{-2}$ and $J_{\rm K}=0.18$, $J_{\rm H}=4.2\times10^{-2}$. In both cases, the impurity entropy directly drops from $4\ln2$ to $\ln\frac{1+\sqrt{5}}{2}$, a characteristic value for the 3CK effect \cite{Tsvelick1985,Affleck1993}. The effective moment correspondingly drops to zero, indicating full screening of the spin degrees of freedom. Thus, the 3CK regime appears in the parameter region where the RKKY interaction is relatively weak and the Kondo impurities may be viewed to be first individually screened by the conduction bath to form three local screened states. These local states further compete to screen the probe impurity via $J_{\rm H}$, causing the 3CK physics. However, since the local states arise from the same shared bath and cannot be fully separated, their mixture will eventually destroy the 3CK fixed point at lower temperatures. We see that the entropy is fully quenched for $J_{\rm K}=0.20$ and $J_{\rm H}=5\times10^{-2}$ but drops to the 2CK value for $J_{\rm K}=0.18$ and $J_{\rm H}=4.2\times10^{-2}$ at lower temperatures, indicating that the 3CK fixed point is indeed unstable and flows to the 2CK or Fermi liquid fixed points. An illustration of this multi-stage, multi-channel process is shown in the inset of Fig.~\ref{fig4}. The role of the probe impurity is supported by comparison with the results for $J_{\rm H}=0$ at $J_{\rm K}=0.20$, where the impurity entropy and the effective moment drop to $\ln2$ and $1/4$, respectively, for a fully decoupled spin down to the lowest temperature. 

The appearance of the 2CK and 3CK regimes suggests that the Kondo screened states of the 3IK model contains a helicity structure, which is detected by the probe impurity through the following effective low-energy Hamiltonian:
\begin{equation}
H_{\rm eff}=J_0\boldsymbol{S}_{4}\cdot\tilde{\boldsymbol{s}}_{00}+J_{\rm E}\boldsymbol{S}_{4}\cdot\left(\tilde{\boldsymbol{s}}_{++}+\tilde{\boldsymbol{s}}_{--}\right),
\end{equation}
where $\tilde{\boldsymbol{s}}_{hh}$ are the spin density operators of the screened states in helicity (auxiliary-bath) basis and $J_0$ and $J_{\rm E}$ are their effective Kondo couplings. Unlike independent bath models \cite{Konig2021,Ferrero2007}, the $h=0$ ($p=0$) channel of the shared bath generally differs from the $h=\pm1$ ($p=1$ and 2) channels, as shown in Fig.~\ref{fig1}(a). As a result, the 4IK model studied here flows to a stable 2CK fixed point if the $J_E$ term dominates and a fully-screened Fermi liquid fixed point if the $J_0$ term dominates, while in the intermediate regime where the two are approximately equal, one finds a narrow temperature window of the unstable 3CK fixed point before it eventually flows to one of the two stable fixed points at lower temperatures. Our NRG simulations of the above effective model indeed confirm this simple picture, thereby clarifying the nature of collective screening in the original 3IK model.

\textit{Conclusions.---}
We have proposed a $C_3$-symmetric 4IK model with an additional probe impurity to detect the collective and individual Kondo screening in the original 3IK model. We show that the same auxiliary-bath construction for the 3IK model can be applied to extended models with additional impurities uncoupled with the shared bath. Our AUNRG calculations find emergent global helicity/spin states and multi-stage, multi-channel Kondo regimes that correspond exactly to the different screening states of the 3IK model. This provides further support for our previously proposed concept of collective Kondo screening in multi-impurity Kondo systems. The screened states may contain nontrivial structures of cluster degrees of freedom that can lead to distinctive physical consequences, demonstrating the predictive power of the newly-developed AUNRG method. The application of the same auxiliary-bath construction to the proposed 4IK model implies that one may design a series of extended models and explore their emergent degrees of freedom and rich many-body phenomena, which makes the multi-impurity Kondo systems a promising platform for in-depth numerical and experimental studies.

This work was supported by the National Natural Science Foundation of China (Grants No. 12404168, No. 12304174, No. 12474140, and No. 12547101) and the National Key R\&D Program of China (Grants No. 2024YFA1408602 and No. 2022YFA1402203).

\end{document}